\documentclass[aps,prb,twocolumn,10pt,longbibliography]{revtex4-1}
\usepackage{amsmath}
\usepackage{amsfonts}
\usepackage{amssymb}
\usepackage{mathtools}
\usepackage[colorlinks=true,citecolor=blue,linkcolor=blue,urlcolor=blue]{hyperref}
\usepackage{bm}
\usepackage{times}
\usepackage{cases}
\usepackage{wasysym}
\usepackage[version=4]{mhchem}
\usepackage{graphicx}
\usepackage[caption=false]{subfig}
 
\begin{document}
\title{Coulombic antiferromagnet in spin-1/2 pyrochlores with dipole-octupole doublets} 
\author{Gang Chen$^{1,2,3}$}
\affiliation{$^{1}$International Center for Quantum Materials, School of Physics, Peking University, Beijing 100871, China}
\affiliation{$^{2}$Department of Physics and HKU-UCAS Joint Institute for Theoretical and Computational Physics at Hong Kong, 
The University of Hong Kong, Hong Kong, China}
\affiliation{$^{3}$Collaborative Innovation Center of Quantum Matter, Beijing 100871, China}

\date{\today}
  
\begin{abstract}
Exotic state of matter could coexist with conventional orders such that the gauge fields 
and fractionalized excitations could prevail in a seemingly ordered state.  
We explore the Coulombic antiferromagnet in the spin-1/2 pyrochlores with dipole-octupole doublets.
This fractionalized antiferromagnetic state carries both antiferromagnetic order and emergent quantum 
electrodynamics with the gapless gauge photon and fractionalized quasiparticles. 
We explain the characteristic physical properties including the thermodynamics and the dynamics of this exotic state. 
This includes for example the anomalously large $T^3$ specific heat and the broad spinon continuum 
in the inelastic neutron scattering.  We discuss the experimental and material's relevance and 
expect this work to inspire interests in the search of exotic physics among the ordered magnets. 
\end{abstract}

\maketitle



Exotic quantum states of matter with long-range quantum entanglement 
such as fractional quantum Hall effects and quantum spin liquids 
are characterized by emergent gauge fields and fractionalized excitations~\cite{Xiao:803748}. 
Due to the emergent non-local gauge structures, these states are usually robust against 
weak local perturbations. Moreover, the emergent gauge field and fractionalization 
could survive even in the presence of long-range orders. This happens, for example when 
the residual interactions between the fractionalized quasiparticles induce the condensation 
of composite objects without completely destroying the internal gauge structures. 
Such a scenario was theoretically proposed about twenty years ago for the superconducting cuprates 
where the hosting exotic state was suggested to be $\mathbb{Z}_2$ topological 
order~\cite{PhysRevB.62.7850,PhysRevB.60.1654,2003IJMPB..17.2821L,PhysRevB.64.014518}. 
Over there, the bilinear of the fermionic spinons was condensed to generate 
the antiferromagnetic long-range order while the fractionalization and 
the $\mathbb{Z}_2$ gauge structure persist. To distinguish it from the conventional 
antiferromagnet, this fractionalized antiferromagnet was dubbed ``AF$^{\ast}$ state''. 
Although its connection to the cuprates remains illusive, such an exotically ordered state 
points to the important possibility of emergent exotic physics in the seemingly ordered 
systems, i.e. the coexistence of long-range order with the long-range quantum
entanglement.

\begin{figure}[b] 
	\centering
	\includegraphics[width=8.5cm]{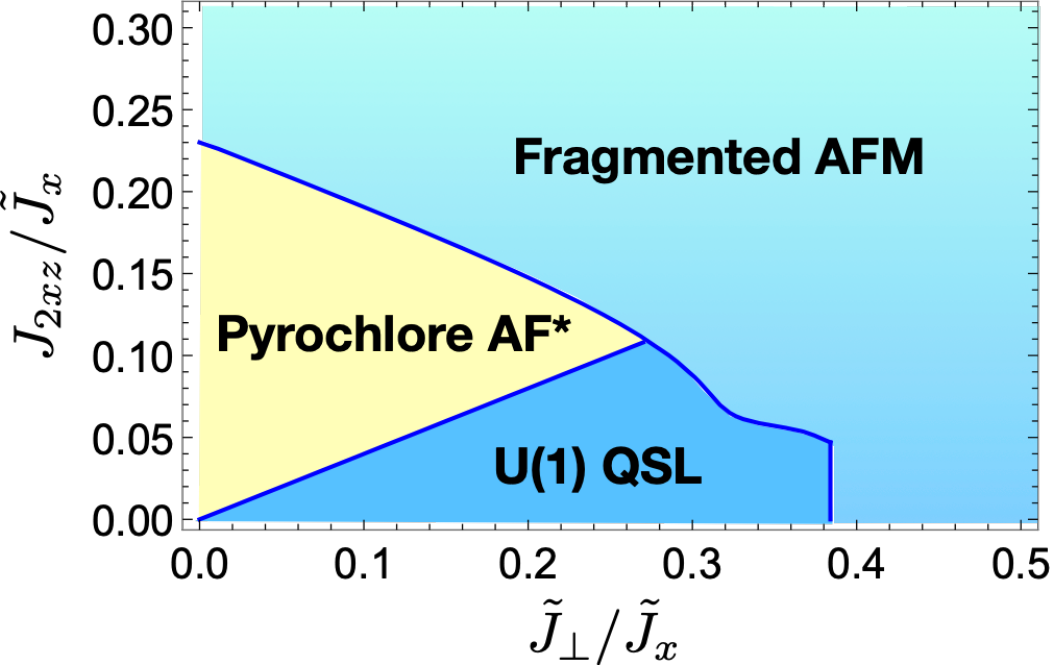}
	\caption{(Color online.) Phase diagram for $H_a$ in Eq.~\eqref{eq3} from the gauge mean-field theory.
	The phase boundary between the pyrochlore AF$^{\ast}$ state and the fragmented AFM  
	is continuous via the spinon condensation. The remaining phase boundaries are all first order. 
	U(1) QSL refers to the pyrochlore U(1) spin liquid. The AF$^{\ast}$ state is referred as the 
	Coulombic antiferromagnet. }
	\label{fig1}
\end{figure}

Despite the possibility of exoticity in the long-range ordered systems, 
the popular search of exotic physics is to examine spin liquid candidates 
among disordered magnets with frustration such that the frustration 
enhances the quantum fluctuations and suppresses the magnetic    
orders down to zero temperature~\cite{Savary_2016,Knolle_2019}. 
In this Letter, we consider the possibility of fractionalized antiferromagnet
in the spin-1/2 quantum pyrochlore antiferromagnet with dipole-octupole doublets
where the emergent U(1) gauge field and the fractionalization coexist 
with the antiferromagnetic order. The pyrochlore magnets have been 
an active topic in the modern research of quantum magnetism and 
have attracted quite some attention in last two decades~\cite{RevModPhys.82.53,Rau_2019}. 
The classical spin ice has been observed in the pyrochlore spin ice 
magnets and understood based on the interacting Ising spins~\cite{Bramwell_2001,Castelnovo_2008,Gingras_2009,2008Natur.451...22T}.
The quantum counterpart, referred as pyrochlore quantum spin ice 
or pyrochlore ice U(1) spin liquid, 
is less conclusive~\cite{PhysRevB.69.064404,Gingras_2014,Savary_2016,PhysRevB.86.104412,Bramwell_2020}. 
There have been two kinds of relevant models~\cite{Savary_2016,PhysRevX.1.021002,PhysRevLett.112.167203,  
PhysRevB.86.104412,PhysRevLett.108.037202,PhysRevB.83.094411, 
Gingras_2014}. Due to the low energy scales of 
the emergent excitations such as the gauge photons and the fractionalized 
excitations, the experimental side is less conclusive compared to the classical case. 
So far, the existing candidates that remain disordered and promising 
are the Tb-based~\cite{PhysRevLett.98.157204}, 
Pr-based pyrochlores~\cite{cmab8423bib105,cmab8423bib106} 
with the non-Kramers doublets, and more recently 
the Ce-based pyrochlores, Ce$_2$Sn$_2$O$_7$ and Ce$_2$Zr$_2$O$_7$,  
with the dipole-octupole (DO) doublets~\cite{PhysRevLett.115.097202,PhysRevLett.122.187201,Gao_2019,
PhysRevB.95.041106,PhysRevResearch.2.013334,Sibille_2020,PhysRevX.12.021015,Bhardwaj_2022,PhysRevResearch.2.023253,chen2023distinguishing}. 
No such fractionalized antiferromagnet was previously obtained 
for the conventional Kramers doublets like Yb$^{3+}$ ions or the 
non-Kramers doublets like Pr$^{3+}$, and thus we turn to the dipole-octupole doublets
and explore its properties and the possible existence. We show the system could 
stabilize another exotic quantum state, namely, the Coulombic antiferromagnet,
in addition to the pyrochlore U(1) spin liquid. This fractionalized antiferromagnet 
is analogous to the AF$^{\ast}$ state for the cuprates except that the gauge sector 
is U(1) in 3D and all the fractionalized quasiparticles are gapped.
Thus, we refer this exotic state as ``pyrochlore AF$^{\ast}$ state''.

To begin with, we consider the generic symmetry-allowed spin model for the DO doublets with
\begin{eqnarray}
H &=& \sum_{\langle ij \rangle } \Big[  J_x S^x_i S^x_j+J_z S^z_i S^z_j + J_{xz} (S^x_i S^z_j + S^z_i S^x_j) 
\nonumber \\
&& \quad\quad +  J_y S^y_i S^y_j  \Big] 
  -  \sum_{\langle\langle ij \rangle\rangle} J_2 S^z_i S^z_j + \cdots,
\label{eq1}
\end{eqnarray}
where the spin ${\boldsymbol S}_i$ is defined 
on the DO doublet and $\langle ij \rangle$ ($\langle\langle ij \rangle\rangle$) 
refers to the nearest (second) neighbors.
The terms inside ``$[ \,\,]$'' are the 
symmetry-allowed interactions between the nearest neighbors,
while ``$\cdots$'' refers to other interactions such as the dipole-dipole 
interaction between the $S^z$-component  
and the superexchange beyond the nearest neighbors. 
For the unfrustrated regime of the nearest-neighbor model, 
the ground state can be well-understood, and no fractionalized antiferromagnet (AF$^{\ast}$) 
phase was found~\cite{PhysRevLett.112.167203,PhysRevB.95.041106}. 
Thus, extra interactions and/or the frustrated regime are required to support its existence,
and we have included a second-neighbor $S^z$-$S^z$ interaction in our study.

Microscopically~\cite{PhysRevLett.112.167203}, the $S^z$ ($S^x$ or $S^y$) 
component for the DO doublet carries the magnetic dipole (octupole) moment. 
One should keep the $S^z$-$S^z$ coupling if the dipole-dipole interaction 
is considered. From the symmetry analysis, however, $S^z$ and $S^x$ ($S^y$) 
transform identically as a magnetic dipole (octupole) under the space group. 
This is the symmetry reason why there exists the $S^x$-$S^z$ coupling in Eq.~\eqref{eq1}. 
If one restricts to the nearest-neighbor model, one can apply a rotation about 
the $y$-axis to eliminate the crossing term between $S^x$ and $S^z$. 
With the further-neighbor interactions, it becomes impossible because such a 
rotation immediately re-generates the crossing terms from the further neighbors. 
After the rotation, Eq.~\eqref{eq1} takes a new form, 
\begin{eqnarray}
\label{eq2}
H &=& \sum_{\langle ij \rangle } 
\Big[  \tilde{J}_x \tilde{S}^x_i \tilde{S}^x_j 
+ \tilde{J}_z \tilde{S}^z_i \tilde{S}^z_j 
+  J_y S^y_i S^y_j  \Big] 
\nonumber \\
&-&\sum_{\langle\langle ij \rangle\rangle } J_2
 \Big[\cos^2 \theta {\tilde S}^z_i {\tilde S}^z_j  
 + \sin^2 \theta {\tilde S}^x_i {\tilde S}^x_j 
 \nonumber \\
 && \quad\quad\quad
+ \cos \theta \sin \theta ( {\tilde S}^x_i {\tilde S}^z_j + {\tilde S}^z_i {\tilde S}^x_j  ) 
\Big] + \cdots,
\end{eqnarray}
where ${\tilde{S}^x_i = \cos \theta S_i^x + \sin \theta S_i^z}$, ${\tilde{S}^z_i = \cos \theta S_i^z - \sin \theta S_i^x}$,
$\tilde{J}_x$ and $\tilde{J}_z$ are the rotated couplings and are related to the old ones~\cite{PhysRevLett.115.097202},
and the crossing term reappears. As we show below, the irremovable crossing term is responsible for the emergence 
of the pyrochlore AF$^{\ast}$ state from the U(1) spin liquid. Since other interactions in ``$\cdots$'' necessarily 
renormalize the $J_2$ interactions, to capture the qualitative physics, we here consider an alternative model 
with the renormalized coupings,
\begin{eqnarray}
H_a   &=& 
  \sum_{\langle ij \rangle } \Big[ \tilde{J}_x \tilde{S}^x_i \tilde{S}^x_j
-\tilde{J_{\perp}} (\tilde{S}^z_i \tilde{S}^z_j + S^y_i S^y_j)
 \Big]
\nonumber \\
&-&  \sum_{\langle\langle ij \rangle\rangle}   
J_{2xz} ( {\tilde S}^x_i {\tilde S}^z_j + {\tilde S}^z_i {\tilde S}^x_j  ).
\label{eq3}
\end{eqnarray}
To avoid further complication, we have set ${\tilde{J}_z = {J}_y = - \tilde{J}_{\perp}}$. 
$H_a$ is our minimal model to reveal the pyrochlore AF$^{\ast}$ state.

We start from the large antiferromagnetic $\tilde{J}_x$ regime and 
consider the instability to the nearby phases. With the large antiferromagnetic
$\tilde{J}_x$, the system prefers the ``two-plus two-minus'' ice configuration 
for $\tilde{S}^x$ and realizes a pyrochlore U(1) spin liquid with the perturbed quantum fluctuations.
This is a reasonable starting point because both the pyrochlore U(1) spin liquid
and the AF$^{\ast}$ state share the same gauge structure.
In this limit, the system is characterized by the emergent U(1) gauge 
field and the fractionalized excitations. To reveal it, we implement 
the spinon-gauge construction and express the spin operators 
as~\cite{PhysRevLett.108.037202,PhysRevB.86.104412,PhysRevLett.112.167203}
\begin{eqnarray}
\tilde{S}^{+}_{{\boldsymbol r} , {\boldsymbol r} +{\boldsymbol e}_{\mu}}
&\equiv & \tilde{S}^z_{{\boldsymbol r} , {\boldsymbol r} +{\boldsymbol e}_{\mu}} 
- i S^y_{{\boldsymbol r} , {\boldsymbol r} +{\boldsymbol e}_{\mu}} 
= \Phi^{\dagger}_{\boldsymbol r} \Phi^{}_{{\boldsymbol r}+{\boldsymbol e}_{\mu}} 
\tilde{s}_{{\boldsymbol r} , {\boldsymbol r} +{\boldsymbol e}_{\mu}  }^{+}, 
\label{eq4}
\\
&& \quad\quad
\tilde{S}_{{\boldsymbol r} , {\boldsymbol r} +{\boldsymbol e}_{\mu}  }^x 
= \tilde{s}_{{\boldsymbol r} , {\boldsymbol r} +{\boldsymbol e}_{\mu}  }^x,
\label{eq5}
\end{eqnarray}
where the pyrochlore spin is now interpreted as sitting on the link
connecting the centers ${\boldsymbol r}$ and ${{\boldsymbol r} +{\boldsymbol e}_{\mu} }$
of the neighboring tetrahedra. The tetrahedral centers form a diamond lattice,
and ${\boldsymbol e}_{\mu}$ refers to the four nearest-neigbor vectors that
connect the I sublattice sites to the II sublattice sites. 
Here $\tilde{\boldsymbol s}_{{\boldsymbol r} , {\boldsymbol r} +{\boldsymbol e}_{\mu}  }$ 
is a spin-1/2 variable that corresponds to the emergent U(1) gauge field.  
The spinons carry the emergent gauge charge, and
$\Phi^{\dagger}_{\boldsymbol r}$ ($\Phi^{}_{\boldsymbol r}$) creates (annihilates) a spinon
at the diamond site ${\boldsymbol r}$ such that the spinon particle number $Q_{\boldsymbol r}$
satisfies
\begin{eqnarray}
[ \Phi_{\boldsymbol r}, Q_{{\boldsymbol r}'} ] =  \Phi_{\boldsymbol r} \delta_{ {\boldsymbol r}{\boldsymbol r}'},
\quad\quad
[ \Phi_{\boldsymbol r}^\dagger, Q_{{\boldsymbol r}'} ] =  -\Phi_{\boldsymbol r}^\dagger \delta_{ {\boldsymbol r}{\boldsymbol r}'}.
\end{eqnarray}
As the Hilbert space is enlarged by the spinon-gauge construction, 
the constraint ${{Q}_{\boldsymbol r} = \eta_{\boldsymbol r} \sum_{\mu}
 \tilde{s}^x_{ {\boldsymbol r}, {\boldsymbol r}+\eta_{\boldsymbol r}  {\boldsymbol e}_{\mu}  }}$ 
is imposed. The spinon-gauge construction captures the nature of the pyrochlore U(1) spin liquid 
as a string-net condensed phase~\cite{PhysRevB.71.045110}, where $\tilde{S}^{\pm}_i$ 
corresponds to the shortest open string with the spinons at the two ends. The model $H_a$ 
becomes 
\begin{widetext}
\begin{equation}
H_a  = 
\sum_{{\boldsymbol r} } \frac{\tilde{J}_x}{2}    Q_{\boldsymbol r}^2
- \sum_{{\boldsymbol r} } \sum_{ \mu \neq \nu}{\frac{\tilde{J}_{\perp}}{2}} 
{ {\Phi^\dagger_{{\boldsymbol r} + \eta_{\boldsymbol r} {\boldsymbol e}_{\mu}}} 
\Phi^{}_{{\boldsymbol r}  +\eta_{\boldsymbol r} {\boldsymbol e}_{\nu}  }} 
 \tilde{s}^{-\eta_{\boldsymbol r}}_{{\boldsymbol r}, {\boldsymbol r} + {\eta}_{\boldsymbol r} {\boldsymbol e}_{\mu}} 
 \tilde{s}^{+\eta_{\boldsymbol r}}_{{\boldsymbol r},{\boldsymbol r}  +\eta_{\boldsymbol r} {\boldsymbol e}_{\nu}  }  
-  \sum_{{\boldsymbol r} \in \text{I} } \sum_{\mu} \sum_{ j \in [{\boldsymbol r},{\boldsymbol r} + {\boldsymbol e}_{\mu}]_2}  
  \frac{J_{2xz}}{2}   \big( {\Phi^\dagger_{\boldsymbol r} \Phi_{{\boldsymbol r} + {\boldsymbol e}_{\mu}}^{}  
 \tilde{s}^+_{{\boldsymbol r},{\boldsymbol r} + {\boldsymbol e}_{\mu}}  
+  h.c. } \big)
 \tilde{s}^x_j 
 \label{eq7}
\end{equation}
where $[{\boldsymbol r},{\boldsymbol r} + {\boldsymbol e}_{\mu}]_2$ refers to the 
set of the second neighbors from this site at ${{\boldsymbol r} +{\boldsymbol e}_{\mu}/2}$. 
To solve $H_a$, we decouple it into the spinon sector and the gauge field sector with
\begin{eqnarray}
H_{\text{spinon}} & = &  \sum_{{\boldsymbol r}} \frac{\tilde{J}_x}{2}    Q_{\boldsymbol r}^2
- \sum_{{\boldsymbol r} } \sum_{\mu \neq \nu}{\frac{1 }{2}}\tilde{J}_{\perp} \chi_1 
{\Phi^\dagger_{{\boldsymbol r} + \eta_{\boldsymbol r} {\boldsymbol e}_{\mu} } 
\Phi^{}_{{\boldsymbol r}  +\eta_{\boldsymbol r} {\boldsymbol e}_{\nu}  }} 
 -  \sum_{{\boldsymbol r} \in \text{I} } \sum_{\mu}  
  6 J_{2xz} \chi_2   \big( \Phi^\dagger_{\boldsymbol r} \Phi_{{\boldsymbol r} + {\boldsymbol e}_{\mu}}^{}  
+  \Phi^{\dagger}_{{\boldsymbol r} + {\boldsymbol e}_{\mu}} \Phi^{}_{\boldsymbol r}  
   \big) ,
   \label{eq8}
      \\
   H_{\text{gauge}} &=& - \sum_{\langle ij \rangle} \tilde{J}_{\perp} I_1 ( \tilde{s}^z_i \tilde{s}^z_j +  {s}^y_i {s}^y_j  )
- \sum_{ \langle\langle ij \rangle\rangle }     J_{2xz} I_2 (\tilde{s}^x_i \tilde{s}^z_j +\tilde{s}^z_i \tilde{s}^x_j  )  ,
\end{eqnarray}
\end{widetext}
where we have set 
\begin{eqnarray}
\chi_1 &=& \langle \tilde{s}^+_i \tilde{s}^-_j \rangle , \quad {\text{for }} \langle ij \rangle , \\
\chi_2 &=& \langle \tilde{s}^+_i \tilde{s}^x_j \rangle,  \quad {\text{for }} \langle\langle ij \rangle\rangle ,
\end{eqnarray}
and 
\begin{eqnarray}
I_1 & = & \langle  \Phi^\dagger_{\boldsymbol r} \Phi^{}_{{\boldsymbol r}'}\rangle , \quad {\text{for }}  \langle\langle ij \rangle\rangle , \\
I_2  &=& \langle  \Phi^\dagger_{\boldsymbol r} \Phi^{}_{{\boldsymbol r}'}\rangle, \quad {\text{for }}  \langle ij \rangle.
\end{eqnarray}
 These parameters are solved self-consistently~\cite{SM}. 
Due to the spatial uniformity of the model, these parameters are uniform throughout the system. 
The gauge mean-field phase diagram is depicted in Fig.~\ref{fig1}. 
Two exotic phases, the pyrochlore U(1) spin liquid and the pyrochlore AF$^{\ast}$ state, 
occupy the regions with small $\tilde{J}_{\perp}$ and $J_{2xz}$. 
The transition between these two states in Fig.~\ref{fig1} is first order.
Both states have gapped spinon and gapless U(1) gauge photon excitations,
while the pyrochlore AF$^{\ast}$ state has an antiferromagnetic
all-in all-out order.  The fragmented AFM in Fig.~\ref{fig1} is a 
conventional antiferromagnet with an all-in all-out order, 
and can be obtained from the AF$^{\ast}$ state
via the spinon condensation at the $\Gamma$ point. 
Such a transition is an Anderson-Higgs' transition. 
The visible part of the magnetic order is in the $S^z$ component and
behaves as 
\begin{eqnarray}
\langle S_i^z \rangle &=& \cos \theta \langle \tilde{S}^z_i \rangle + \sin \theta \langle \tilde{S}^x_i \rangle 
\nonumber \\
&=&   \frac{1}{2} \cos \theta \big[ \langle \Phi^\dagger_{\boldsymbol r} \Phi^{}_{{\boldsymbol r}'}
 \rangle \langle\tilde{s}^{\dagger}_{{\boldsymbol r}{\boldsymbol r}'} \rangle  + h.c.\big]
 + \sin \theta \langle\tilde{s}^x_i \rangle.
 \label{order}
\end{eqnarray}
The ``fragmented AFM'' is used to capture this moment fragmentation 
where the ordered momentum is fragmented into the 
gauge link piece and the spinon condensate piece.  
Although the $S^x$ moment also develops order, 
it is invisible or hidden due to its magnetic multipolar nature.
Moreover, the $S^z$ moment, despite being ordered, could simultaneously 
function as a disordering operator to flip the $S^x$ order and generate the 
magnetic excitations.

From Eq.~\eqref{order}, it is clear that the all-in all-out ($S^z$) order appears 
when there exists the spinon tunneling between two diamond sublattices 
and/or a non-vanishing emergent gauge field. Due to the generic coupling between 
$S^x$ and $S^z$ or between $\tilde{S}^x$ and $\tilde{S}^z$, these two conditions 
are actually concomitant. In the above, we have explicitly shown that
the pyrochlore AF$^{\ast}$ state appears from the irremovable coupling 
between $\tilde{S}^x$ and $\tilde{S}^z$ and supports both contributions 
in Eq.~\eqref{order} non-vanishing. It is also possible that, the spinon interaction 
leads to the condensation in the spinon particle-hole channel and results 
in an antiferromagnetic order. The key physical properties of the pyrochlore 
AF$^{\ast}$ state, however, are independent from the physical origin and 
are discussed below. 

\begin{table}[b]
\begin{tabular}{p{2.5cm}p{1cm}p{1cm}p{1cm}p{1cm}p{1cm}}
\hline\hline
\rule{0pt}{15pt}Phases & $\langle \Phi \rangle $ & $\langle \tilde{s}^x \rangle $ & $\langle \tilde{s}^{\pm} \rangle $ &  $\langle S^x \rangle $  & $\langle S^z \rangle $  \\
\rule{0pt}{15pt} U(1) QSL & $=0$ & $=0$ &$\neq 0$ & $=0$ &$=0$ \\
\rule{0pt}{15pt} {AF$^{\ast}$ state} & $=0$ & $\neq 0$ & $ \neq 0 $ & $\neq 0$ & $ \neq 0 $ \\
\rule{0pt}{15pt}{Fragmented AFM} & $ \neq 0 $ & $\neq 0 $ & $\neq 0 $ & $\neq 0 $ & $\neq 0 $ \\
\\
\hline\hline
\end{tabular}
\caption{Description of each phase within gauge mean-field theory.}
\end{table}

\begin{figure}[t] 
	\centering
	\includegraphics[width=7cm]{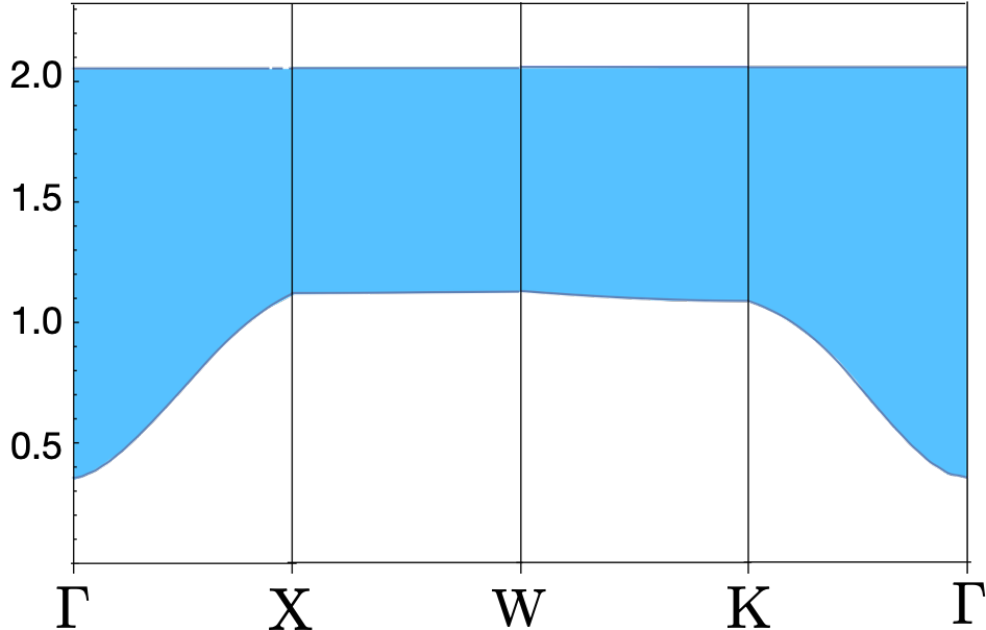}
 	\caption{Spinon continuum along the high symmetry momentum 
	lines 	of the pyrochlore AF$^{\ast}$ state. Here  
	${t_1 = 0.025}, {t_2 =0.02}$, and $\tilde{J}_x$ is set to unity. }
	\label{fig2}
\end{figure}

Despite the presence of the all-in all-out order, the quantum fluctuations of 
the pyrochlore AF$^{\ast}$ state are still governed by the fractionalized spinons 
and the emergent U(1) gauge fluctuations. Thus, the gapless U(1) gauge photon 
is responsible for the $T^3$ specific heat at low temperatures. 
Due to the low energy scale of the exchange coupling, the coefficient 
of the $T^3$ specific heat should be quite large and visible experimentally. 
We further explore the spectroscopic consequences of the pyrochlore AF$^{\ast}$ state. 
Due to the fractionalized nature of this state, the excitation spectrum should be quite
different from the spin-wave excitation for the conventional ordered antiferromagnet.
This can be analyzed from a more phenomenological treatment that introduces 
the all-in all-out order on top of the fractionalized spin liquid state. 
The resulting spinon Hamiltonian will be of an identical 
structure as Eq.~\eqref{eq8}, and we write it down here,
\begin{eqnarray}
H_{\text{AF}^{\ast}} &=&  \sum_{{\boldsymbol r} } \frac{\tilde{J}_x}{2}    
Q_{\boldsymbol r}^2
- \sum_{{\boldsymbol r}  } \sum_{\mu\neq \nu} t_1
{\Phi^\dagger_{{\boldsymbol r} + \eta_{\boldsymbol r} {\boldsymbol e}_{\mu} } 
\Phi^{}_{{\boldsymbol r}  +\eta_{\boldsymbol r} {\boldsymbol e}_{\nu}  }} 
\nonumber 
\\
 && -  \sum_{{\boldsymbol r} \in \text{I} } \sum_{\mu}  
  t_2 \big( \Phi^\dagger_{\boldsymbol r} \Phi_{{\boldsymbol r} + {\boldsymbol e}_{\mu}}^{}  
+  \Phi^{\dagger}_{{\boldsymbol r} + {\boldsymbol e}_{\mu}} \Phi^{}_{\boldsymbol r}  
   \big) ,
   \label{eq11}
\end{eqnarray}
where the inter-sublattice hopping ${t_2 \neq 0}$ when the magnetic order is present. 
This is a bit different from the case without the sublattice mixing where the 
spinon number is separately conserved on each diamond sublattice and 
the two spinon bands are degenerate~\cite{PhysRevB.69.064404}. 
In our case here the two spinon dispersions are not degenerate, 
and we have
\begin{equation}
\omega_{\pm} ({\boldsymbol k}) = \sqrt{2 \tilde{J}_x} \big[ \lambda 
      - t_1 \sum_n \cos ({\boldsymbol k} \cdot {\boldsymbol a}_n) 
 \pm t_2  | \sum_{\mu} e^{i {\boldsymbol k} \cdot {\boldsymbol e}_{\mu}} |
\big]^{\frac{1}{2}} ,
\end{equation}
where $\{ {\boldsymbol a}_n \}$ refers to the twelve second-neighbor vectors 
on the diamond lattice, and $\lambda$ is the Lagrangian multiplier used to fix 
the constraint ${|\Phi^2_{\boldsymbol r}|=1}$ and determined self-consistently. 
Since the ${S}^z$ moment in Eq.~\eqref{order} contains the spinon bilinear 
and is the only component that is linearly coupled to the external magnetic field, 
the spinon-pair excitations are included in the ${S}^z$-${S}^z$ correlation 
and can be detected in the inelastic neutron scattering measurement. 
From the energy-momentum conservation, the spinon-pair excitation is 
characterized with the energy-momentum relation,
\begin{eqnarray}
{\Omega({\boldsymbol q}) = 
\omega_{\mu} ({\boldsymbol k}_1 ) + \omega_{\nu} ({\boldsymbol q} - {\boldsymbol k}_1 )} ,
\end{eqnarray}
where ${\mu,\nu=\pm}$ refer to the spinon branches. 
Due to the freedom of the spinon momentum ${\boldsymbol k}_1$,
the above excitation corresponds to a continuum in both energy 
and momentum domains. In Fig.~\ref{fig2}, we plot the 
momentum-resolved energy range of the two-spinon continuum.

Apart from the gapped spinon continuum and the gapless gauge photon in the dynamics, 
there also exists the electric monopole continuum in the $\tilde{S}^x$-$\tilde{S}^x$ 
correlation for the pyrochlore AF$^{\ast}$ state. 
For the pyrochlore U(1) spin liquid, the electric monopole experiences a background dual
U(1) gauge flux that is a $\pi$ flux in the unit cell, and the spectrum has an enlarged spectral
period in the reciprocal space. In the pyrochlore AF$^{\ast}$ state, 
because ${\langle \tilde{S}^x \rangle \neq 0}$, the background dual 
U(1) gauge flux is deviated from the $\pi$ flux, and the electric monopole 
will have a Hofstadter band structure that can be manifested 
in the monopole continuum~\cite{PhysRevResearch.2.013066}.
As the ${S}^z$-${S}^z$ correlation contains 
both $\tilde{S}^z$-$\tilde{S}^z$ and $\tilde{S}^x$-$\tilde{S}^x$ correlations~\cite{PhysRevB.95.041106}, 
the monopole continuum can be revealed. In contrast to the gapped monopoles and 
the gapless gauge photon that have a close and low energy scale, the spinons appear 
in a much higher energy and may be more accessible in the scattering measurements~\cite{PhysRevB.96.085136}.  
In general, the emergent parton-gauge dynamics for the pyrochlore AF$^{\ast}$ state
is fundamentally different from the simple magnons for a conventionally ordered magnet. As a comparison,  
we further plot the magnon dispersion for the fragmented AFM in the Supplementary material~\cite{SM}. 
Among the four magnon branches, two are flat bands, and the other two are dispersive. 
 
\emph{Discussion.}---As the pyrochlore AF$^{\ast}$ state carries an antiferromagnetic order,
there should exist a finite-temperature phase transition. The antiferromagnetic order 
occurs in the particle-hole channel of the spinons and does not carry any emergent 
gauge charge. Thus, the Ginzburg-Landau theory for such a transition is absent of 
the gauge degrees of freedom, and we expect this transition to be a 3D
Ising transition. As for the material relevance, we propose 
Nd$_2$Sn$_2$O$_7$ as a candidate for the pyrochlore AF$^{\ast}$ state. 
The magnetic transition~\cite{BLOTE1969549} in Nd$_2$Sn$_2$O$_7$ 
was found to be continuous~\cite{PhysRevB.92.144423}. It would be illuminating 
to examine the critical exponents at the transition. 

Below the ordering transition, Nd$_2$Sn$_2$O$_7$ shows an anomalously 
large $T^3$ specific heat, and Ref.~\onlinecite{PhysRevB.92.144423} 
expected a linearly dispersive mode with an excitation velocity ${v_{\text{ex}} = 55}$m/s. 
There are two reasons against the Goldstone magnon mode interpretation for this mode. 
First, the Nd$^{3+}$ ground state doublet 
is a DO doublet~\cite{PhysRevLett.112.167203} and is well separated from 
the excited ones by a crystal field gap of 26meV~\cite{PhysRevB.92.144423} 
such that the effective spin-1/2 moment from the ground states captures   
the magnetic properties below $\sim$20K. The effective model 
does not have any continuous symmetry~\cite{PhysRevLett.112.167203} 
to support gapless Goldstone modes, nor have an accidental continuous 
degeneracy to support pseudo-Goldstone modes via quantum order 
by disorder~\cite{PhysRevLett.109.167201,PhysRevLett.109.077204}.  
Second, how about a fine-tuned spin model that is close to the point 
with a continuous spin symmetry? To obtain the all-in all-out order,
the model should be a ferromagnetic model with a continuous symmetry. 
Little frustration would be expected for the continuous ferromagnetic case,
which is incompatible with the frustrated nature of the slow paramagnetic spin dynamics
from the inelastic neutron and $\mu$SR measurements~\cite{PhysRevB.92.144423,PhysRevB.95.134420}. 
Moreover, the ordered fraction of Nd$^{3+}$ moments is not small~\cite{PhysRevB.92.144423}, 
so it is unlikely to be proximate to a quantum transition with gapless critical modes. 
If the pyrochlore AF$^{\ast}$ state proposal 
for Nd$_2$Sn$_2$O$_7$ is relevant, $v_{\text{ex}}$ would be interpreted as
 the speed of emergent gauge photon. To further examine the proposal, 
we suggest the inelastic neutron scattering measurement 
to directly detect the gauge photon as well as the continuum 
of the spinons and the electric monopoles.

In contrast to Nd$_2$Sn$_2$O$_7$, Nd$_2$Zr$_2$O$_7$ and Nd$_2$Hf$_2$O$_7$ 
are deep in the ``fragmented AFM'' state with an all-in all-out order 
in the $S^z$ component and the well-defined spin-wave 
modes~\cite{PhysRevLett.124.097203,PhysRevB.92.224430,PhysRevB.92.184418,Petit2016,PhysRevB.94.104430}. 
As the actual order involves the hidden $S^x$ component, the $S^z$ operator is then 
responsible for flipping the order in $S^x$ in addition to having the order within itself. 
This interesting moment fragmentation~\cite{PhysRevX.4.011007} 
arises from the $J_{xz}$ crossing term~\cite{PhysRevLett.112.167203} 
in Eq.~\eqref{eq1}, and has been well-understood from 
the peculiar microscopic properties and the model 
for the DO doublets of the Nd$^{3+}$ ions~\cite{PhysRevLett.124.097203,Petit2016,PhysRevB.94.104430}. 
Here we do not get into too much details. Another 
Nd-compound Nd$_2$GaSbO$_7$ with the all-in all-out order 
was experimentally studied and the moment fragmentation physics 
was absent~\cite{PhysRevB.103.214419}. 
The continuum-like feature in the ordered regime
seems more compatible with the continuous excitations of the 
AF$^{\ast}$ state. Due to the intrinsic disorder, however,
more information needs to be collected and examined.

\emph{Acknowledgments.}---This work is supported by the Ministry of Science and Technology of China with Grants 
No.~2021YFA1400300, the National Science Foundation of 
China with Grant No.~92065203, and by the Research Grants Council of Hong Kong with C7012-21GF. 

\bibliography{ref.bib}

\appendix

\section{Conventions for the coordinates}
\label{sec1}

Here we list the conventions for the coordinates for the main text and this Supplementary material. 
The pyrochlore lattice has four sublattices, and the coodinates of the reference points from each 
sublattice are given as
\begin{eqnarray}
&& \text{0th sublattice}: \quad {\boldsymbol b}_0 = (0,0,0), \\
&& \text{1st sublattice}: \quad \,  {\boldsymbol b}_1 = (0,\frac{1}{4},\frac{1}{4}), \\
&& \text{2nd sublattice}: \quad  {\boldsymbol b}_2 = (\frac{1}{4},0,\frac{1}{4}), \\ 
&& \text{3rd sublattice}: \quad \,  {\boldsymbol b}_3 = (\frac{1}{4},\frac{1}{4},0). 
\end{eqnarray}
For each site on the pyrochlore lattice, there are twelve second-neighbor sites, 
and the connecting vectors are different for different sublattices. This information 
is used in the calculation of the gauge sector of the mean-field theory
and the Weiss mean-field theory in Sec.~\ref{sec2}. 

The centers of the tetrahedra on the pyrochlore lattice form a diamond lattice. 
This is where the spinon resides. The reference points for the two diamond 
sublattices are 
\begin{eqnarray}
&& \text{I sublattice}: \quad \, ({+\frac{1}{8},+\frac{1}{8},+\frac{1}{8}} ) , \\
&&  \text{II sublattice}: \quad \, ({-\frac{1}{8},-\frac{1}{8},-\frac{1}{8}}) . 
\end{eqnarray}

The four nearest-neighbor vectors on the diamond lattice are given as 
\begin{eqnarray}
&& {\boldsymbol e}_0 = (+\frac{1}{4},+\frac{1}{4},+\frac{1}{4}), \\
&& {\boldsymbol e}_1 = (+\frac{1}{4},-\frac{1}{4},-\frac{1}{4}), \\
&& {\boldsymbol e}_2 =(-\frac{1}{4},+\frac{1}{4},-\frac{1}{4}), \\
&& {\boldsymbol e}_3 = (-\frac{1}{4},-\frac{1}{4},+\frac{1}{4}),
\label{diamondNN}
\end{eqnarray}
where the subindices correspond to the sublattices of the pyrochlore lattice 
whose positions are the mid-points of the vectors.

\section{Weiss mean-field theory}
\label{sec2}

\begin{figure}[t] 
	\centering
	\includegraphics[width=8.4cm]{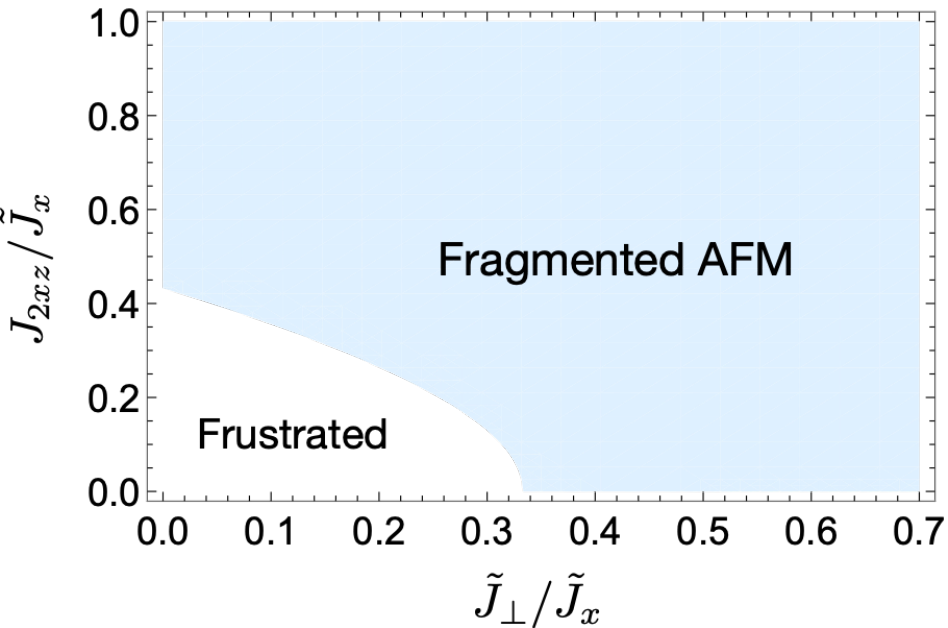}
 	\caption{(Color online.) Phase diagram from the Weiss mean-field theory. 
 	``Frustrated'' refers to the frustrated regime where the Weiss mean-field theory 	fails to give reliable ground states. 
 	``Fragmented AFM'' is identical to the ``Fragmented AFM'' in Fig.~1 of the main text.  
	}
	\label{fig3}
\end{figure}

The main text is using the exotic spinon-gauge construction and treatment 
because we are dealing with exotic quantum phases of matter. In the phase diagram 
that we found there exist the conventional ordered phase like the 
fragmented antiferromagnet. The conventional ordered phases can 
be understood from the more conventional means. 
Thus, we here adopt the conventional Weiss type of mean-field approach to explore the ordered phases. 
Although the Weiss mean-field theory fails to capture the properties of the exotic states, 
it is helpful to ensure the ground for the fully ordered phases in the phase diagram. 
For this purpose, we replace the spin operator as the mean-field order and 
consider the spin as a classical vector. The classical mean-field ground state
is obtained by optimizing the energy of the Hamiltonian in the classical limit,
\begin{eqnarray}
H = \sum_{\boldsymbol k} \sum_{ \mu \nu} \sum_{\alpha\beta} S^{\alpha}_{{\boldsymbol k},\mu} M_{\mu\nu,\alpha\beta}^{} 
({\boldsymbol k}) S^{\beta}_{-{\boldsymbol k},\nu},
\end{eqnarray}
where $M ({\boldsymbol k})$ is the $12\times 12$ ``exchange matrix'' in the momentum space, 
and $S^{\mu}_{\boldsymbol k}$ is the Fourier component of the spin vector with
\begin{eqnarray}
S^{\alpha}_{i} = \frac{1}{\sqrt{N}}\sum_{\boldsymbol k} S^{\alpha}_{{\boldsymbol k},\mu} e^{i {\boldsymbol k} \cdot {\boldsymbol R}_i}.
\end{eqnarray}
Here ${\boldsymbol R}_i$ is the coordinate of the site $i$, $i$ belongs to the $\mu$-th sublattice, $N$ is the 
total number of unit cell. The exchange matrix is given as
\begin{eqnarray}
M  ({\boldsymbol k}) = 
\left[
\begin{array}{ llll }
T_{00}(\boldsymbol k) & T_{01} (\boldsymbol k) & T_{02} (\boldsymbol k) & T_{03} (\boldsymbol k) \\
T_{10}(\boldsymbol k) & T_{11}(\boldsymbol k) & T_{12}(\boldsymbol k) & T_{13}(\boldsymbol k) \\
T_{20}(\boldsymbol k) & T_{21}(\boldsymbol k) & T_{22}(\boldsymbol k) & T_{23}(\boldsymbol k) \\
T_{30}(\boldsymbol k) & T_{31}(\boldsymbol k) & T_{32}(\boldsymbol k) & T_{33}(\boldsymbol k) 
\end{array}
\right],
\end{eqnarray}
where $T_{\mu\nu} ({\boldsymbol k})$ is the $3\times 3$ exchange 
matrix between the $\mu$-th and the $\nu$-th sublattices. 
As there does not exist the coupling from the same sublattice, 
the diagonal matrices are all 0 with
\begin{eqnarray}
T_{\mu\mu} (\boldsymbol k) & = & 0.
\end{eqnarray}
For the off-diagonal ones with ${\mu \neq \nu}$, the contribution from the nearest-neighbor interaction
is simply given as 
\begin{eqnarray}
 \cos [{{\boldsymbol k} \cdot {\boldsymbol b}_{\mu}-{\boldsymbol k}  \cdot {\boldsymbol b}_{\nu}} ]  
 \left[ 
\begin{array}{ccc}
\tilde{J}_x  & 0 & 0 \\
0 & J_y & 0 \\
0 & 0 & \tilde{J}_z
\end{array}
\right].
\end{eqnarray}
The contribution from the second neighbor interaction is more complex. There exist 12 
second neighbors for each site and they belong to different sublattices.  
After a careful listing, one obtains for the interaction 
$\sum_{\langle \langle ij \rangle \rangle} -J_{2xz} ( \tilde{S}^x_i \tilde{S}^z_j + \tilde{S}^z_i \tilde{S}^x_j ) $ that,
the contribution to $T_{01} ({\boldsymbol k})$ is 
\begin{eqnarray}
2 \cos \frac{k_x}{2} \cos \big( \frac{k_y - k_z}{4} \big)
 \left[ 
\begin{array}{ccc}
0  & 0 & -J_{2xz} \\
0 & 0 & 0 \\
-J_{2xz} & 0 & 0
\end{array}
\right],
\end{eqnarray}
the contribution to $T_{02} ({\boldsymbol k})$ is
\begin{eqnarray}
2 \cos \frac{k_y}{2} \cos \big( \frac{k_x - k_z}{4} \big)
 \left[ 
\begin{array}{ccc}
0  & 0 & -J_{2xz} \\
0 & 0 & 0 \\
-J_{2xz} & 0 & 0
\end{array}
\right],
\end{eqnarray} 
the contribution to $T_{03} ({\boldsymbol k})$ is
\begin{eqnarray}
2 \cos \frac{k_z}{2} \cos \big( \frac{k_x - k_y}{4} \big)
 \left[ 
\begin{array}{ccc}
0  & 0 & -J_{2xz} \\
0 & 0 & 0 \\
-J_{2xz} & 0 & 0
\end{array}
\right],
\end{eqnarray}
the contribution to $T_{12} ({\boldsymbol k})$ is 
\begin{eqnarray}
2 \cos \frac{k_z}{2} \cos \big( \frac{k_x + k_y}{4} \big)
 \left[ 
\begin{array}{ccc}
0  & 0 & -J_{2xz} \\
0 & 0 & 0 \\
-J_{2xz} & 0 & 0
\end{array}
\right],
\end{eqnarray}
the contribution to $T_{13} ({\boldsymbol k})$ is 
\begin{eqnarray}
2 \cos \frac{k_y}{2} \cos \big( \frac{k_x + k_z}{4} \big)
 \left[ 
\begin{array}{ccc}
0  & 0 & -J_{2xz} \\
0 & 0 & 0 \\
-J_{2xz} & 0 & 0
\end{array}
\right],
\end{eqnarray}
and the contribution to $T_{23} ({\boldsymbol k})$ is 
\begin{eqnarray}
2 \cos \frac{k_x}{2} \cos \big( \frac{k_y + k_z}{4} \big)
 \left[ 
\begin{array}{ccc}
0  & 0 & -J_{2xz} \\
0 & 0 & 0 \\
-J_{2xz} & 0 & 0
\end{array}
\right].
\end{eqnarray}
By setting ${\tilde{J}_z = J_y = - \tilde{J}_{\perp}}$, we solve for the 
ground state of $H_a$ in Eq.~3 of the main text with the Weiss mean-field theory.
The phase diagram is depicted in Fig.~\ref{fig3}. The Weiss mean-field theory
should capture the physics of the ordered phase. As it is expected, with
large $\tilde{J}_{\perp}$ and $J_{2xz}$, the ground state is the fragmented AFM. 
This qualitative result is actually consistent with the result from the gauge mean-field
theory in Fig.~1.

In the fragmented AFM, both $\tilde{S}^x$ and $\tilde{S}^z$ components develop
uniform magnetic orders. The visible part of the order is the $S^z$ component that is
a linear combination of $\tilde{S}^x$ and $\tilde{S}^z$. As the $z$ axis is defined 
locally on each sublattice, although the $\langle S^z \rangle $ is uniform, the overall state
is an antiferromagnetic state. Moreover, as some portion of the ordered moment remains 
in the hidden octupolar part~\cite{PhysRevX.4.011007,Petit2016,PhysRevLett.112.167203}, 
this antiferromagnetic ordered state is referred 
as the ``fragmented AFM''.

In this fragmented AFM, the reason for the presence of non-vanishing $\tilde{S}^x$ and $\tilde{S}^z$ 
order is due to the crossing coupling $J_{2xz}$. The $\tilde{S}^z$ order, works as an effective and 
uniform Zeeman coupling to the $\tilde{S}^x$ moment through the crossing coupling $J_{2xz}$. 
Once $\tilde{S}^z$ develops an uniform order (for example from
the $J_{\perp}$ coupling), it immediately polarizes $\tilde{S}^x$ and generates a $\tilde{S}^x$ order even 
when there exists a nearest-neighbor antiferromagnetic $\tilde{S}^x$-$\tilde{S}^x$ interaction. 


\section{Calculation for gauge mean-field theory}

We sketch the calculation of the phase diagram and the spinon excitation within the
gauge mean-field theory. We first consider the gauge field sector because it is much 
simpler than the spinon sector. The gauge field mean-field Hamiltonian is given as
\begin{eqnarray}
H_{\text{gauge}} &=& -\sum_{\langle ij \rangle } \tilde{J}_{\perp} I_1 ( \tilde{s}^z_i \tilde{s}^z_j + s^y_i s^y_j) 
\nonumber \\
&& - \sum_{\langle\langle ij \rangle\rangle } J_{2xz} I_2 
( \tilde{s}^x_i \tilde{s}^z_j + \tilde{s}^z_i \tilde{s}^x_j) .
\label{eqg}
\end{eqnarray}
It is clear from the spinon mean-field Hamiltonian $H_{\text{spinon}}$, both $I_1$ and $I_2$
are positive. Thus, this ferromagnetic like interaction in Eq.~\eqref{eqg} is quite 
straightforward to solve with a simple classical treatment. We choose a mean-field ansatz with
\begin{eqnarray}
\langle s^z_i \rangle &=& \frac{1}{2} \cos \phi , \\
\langle s^x_i \rangle &=& \frac{1}{2} \sin \phi,
\end{eqnarray} 
where $\phi$ should be determined. The energy of $H_{\text{gauge}}$ is optimized by $\phi$
that satisfies
\begin{eqnarray}
\tan (2\phi) = \frac{4 I_2 J_{2xz}}{I_1 \tilde{J}_{\perp}}. 
\end{eqnarray}
The gauge mean-field parameters, $\chi_1$ and $\chi_2$, are then obtained as 
\begin{eqnarray}
\chi_1 &=& \langle \tilde{s}^+_i \tilde{s}^-_j \rangle =  \langle \tilde{s}^+_i  \rangle \langle \tilde{s}^-_j \rangle  =  \frac{1}{4} \cos^2 \phi , \\
\chi_2 &=& \langle \tilde{s}^+_i \tilde{s}^x_j \rangle = \langle \tilde{s}^+_i \rangle \langle \tilde{s}^x_j \rangle  = \frac{1}{4} \cos \phi \sin \phi . 
\end{eqnarray}
Due to the ferromagnetic interaction in $H_{\text{gauge}}$, the mean-field parameters, $\chi_1$ and $\chi_2$,
are uniform throughout the lattice. 

\begin{figure}[t] 
	\centering
	\includegraphics[width=7cm]{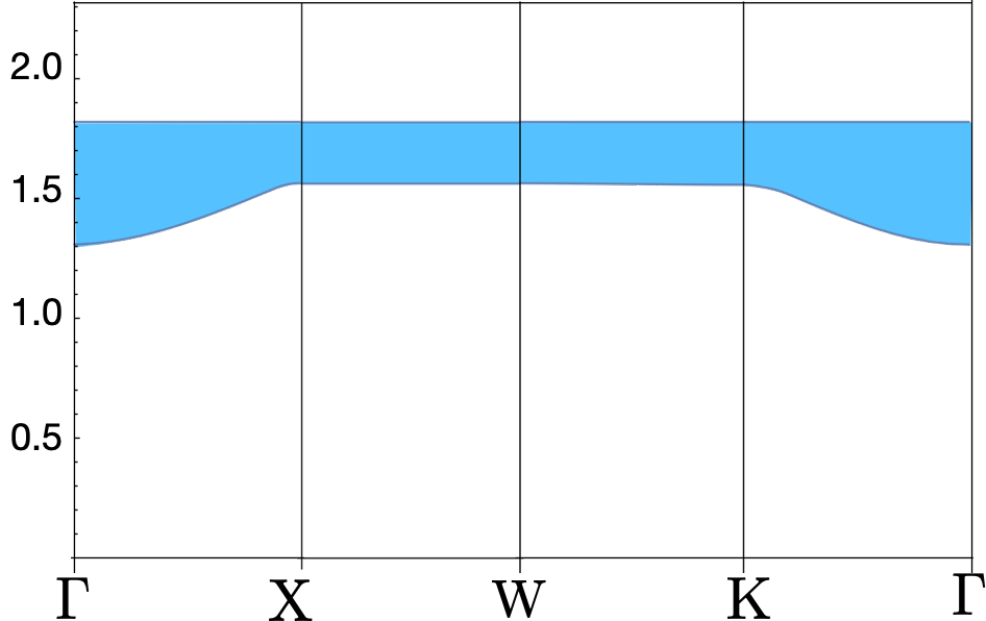}
 	\caption{Spinon continuum along the high symmetry momentum 
	lines 	of the pyrochlore U(1) spin liquid. Here  
	${t_1 = 0.025}, {t_2 =0}$, and $\tilde{J}_x$ is set to unity. 
	The spinon dispersion is obtained for the spinon Hamiltonian in Eq.~11 of the main text.
	}
	\label{fig4}
\end{figure}

Equipped with the above results, one can proceed to solve for the spinon sector. 
The spinon sector Hamiltonian is given as
\begin{eqnarray}
H_{\text{spinon}} & = &  \sum_{{\boldsymbol r}} \frac{\tilde{J}_x}{2}    Q_{\boldsymbol r}^2
- \sum_{{\boldsymbol r} } \sum_{\mu \neq \nu}{\frac{1 }{2}}\tilde{J}_{\perp} \chi_1 
{\Phi^\dagger_{{\boldsymbol r} + \eta_{\boldsymbol r} {\boldsymbol e}_{\mu} } 
\Phi^{}_{{\boldsymbol r}  +\eta_{\boldsymbol r} {\boldsymbol e}_{\nu}  }} 
\nonumber \\
&&
 -  \sum_{{\boldsymbol r} \in \text{I} } \sum_{\mu}  
  6 J_{2xz} \chi_2   \big( \Phi^\dagger_{\boldsymbol r} \Phi_{{\boldsymbol r} + {\boldsymbol e}_{\mu}}^{}  
+  \Phi^{\dagger}_{{\boldsymbol r} + {\boldsymbol e}_{\mu}} \Phi^{}_{\boldsymbol r}  
   \big) .
\end{eqnarray}
It was found to be more convenient to use a rotor representation for the spinon operator
such that ${\Phi_{\boldsymbol r } = e^{-i\varphi_{\boldsymbol r}}}$ and ${|\Phi_{\boldsymbol r} |=1}$.
This approximation neglects the amplitude fluctuation of the spinon fields and should be 
reasonable in the regimes where the spinon matter drives the physics. 

To implement the constraint, we introduce a Lagrangian multiplier $\lambda$ such that
\begin{eqnarray}
H_{\text{spinon}} \rightarrow H_{\text{spinon}}  + \sum_{\boldsymbol r} \lambda_{\boldsymbol r} 
( \Phi_{\boldsymbol r}^\dagger \Phi^{}_{\boldsymbol r} -1),
\end{eqnarray}
where $ \lambda_{\boldsymbol r}$ should be determined self-consistently from the modified spinon Hamiltonian 
and satisfy
\begin{eqnarray}
\langle \Phi_{\boldsymbol r}^\dagger \Phi^{}_{\boldsymbol r}  \rangle  =1
\end{eqnarray}
on every site. 
It is now more convenient to use path integral formulation to solve the spinon sector. 
In a coherent state path integral, we integrate out the ``momentum-like'' field $Q_{\boldsymbol r}$ 
and establish the partition function as
\begin{eqnarray}
{\mathcal Z} = \int {\mathcal D} {\Phi}^\dagger {\mathcal D} \Phi^{}  {\mathcal D} \lambda e^{- {\mathcal S} } e^{-\sum_{\boldsymbol r} \int d\tau \lambda_{\boldsymbol r} (   \Phi_{\boldsymbol r}^\dagger \Phi^{}_{\boldsymbol r}  -1 )}.
\label{partition}
\end{eqnarray}
Here the effective action ${\mathcal S}$ has the form as 
\begin{eqnarray}
{\mathcal S} &=& \int_{d\tau} \sum_{\boldsymbol r} \frac{| \partial_{\tau} \Phi_{\boldsymbol r}  |^2 }{2 \tilde{J}_x} 
- \sum_{{\boldsymbol r} } \sum_{\mu \neq \nu}{\frac{1 }{2}}\tilde{J}_{\perp} \chi_1 
{\Phi^\dagger_{{\boldsymbol r} + \eta_{\boldsymbol r} {\boldsymbol e}_{\mu} } 
\Phi^{}_{{\boldsymbol r}  +\eta_{\boldsymbol r} {\boldsymbol e}_{\nu}  }} 
\nonumber \\
&& -  \sum_{{\boldsymbol r} \in \text{I} } \sum_{\mu}  
  6 J_{2xz} \chi_2   \big( \Phi^\dagger_{\boldsymbol r} \Phi_{{\boldsymbol r} + {\boldsymbol e}_{\mu}}^{}  
+  \Phi^{\dagger}_{{\boldsymbol r} + {\boldsymbol e}_{\mu}} \Phi^{}_{\boldsymbol r}  
   \big). 
   \label{action}
\end{eqnarray}
In a saddle point approximation, we have $\lambda_{\boldsymbol r} = \lambda$. 
This is also demanded by the uniform state requirement. 
 
 The spinon dispersion can be immediately read off from Eq.~\eqref{action} and 
 Eq.~\eqref{partition}, and are given as 
 \begin{eqnarray}
 \omega_{\pm} ({\boldsymbol k})& =& \sqrt{2\tilde{J}_x} \Big[ \lambda -  \frac{\tilde{J}_{\perp} \chi_1}{2} \sum_{\{  {\boldsymbol a}_n \}}  \cos ({\boldsymbol k} \cdot {\boldsymbol a}_n)  
 \nonumber \\
 && \quad\quad \pm 6 J_{2xz} \chi_2 \big| \sum_{ \{ {\boldsymbol e}_{\mu} \}} e^{i {\boldsymbol k}\cdot {\boldsymbol e}_{\mu}} \big|
 \Big]^{1/2},
 \end{eqnarray}
where $\{ {\boldsymbol e}_{\mu} \}$ ($\{ {\boldsymbol a}_n\}$) refers to the set of 4 first-neighbor (12 second-neighbor) vectors on the diamond lattice. 
In both the pyrochlore U(1) spin liquid and the pyrochlore AF$^{\ast}$ state, the spinons are fully gapped. 
Once it becomes gapless, the spinons will be condensed and generate magnetism of some sorts.

The mean-field parameters, $I_1$ and $I_2$, as well as the spinon amplitude, are the spinon bilinears. It is quite convenient to evaluate from the path integral approach. We find that,
\begin{eqnarray}
\langle  \Phi_{\boldsymbol r}^\dagger \Phi^{}_{\boldsymbol r} \rangle =\frac{1}{2N_{u.c.}} \sum_{\boldsymbol k} 
\tilde{J}_x \big[ \frac{1}{\omega_+ ({\boldsymbol k})} + \frac{1}{\omega_- ({\boldsymbol k})}    \big],
\end{eqnarray}
and this quantity should be equal to 1. Here $N_{u.c.}$ refers to the total number of unit cell. 
Moreover, for the second-neighbor sites $\langle\langle {\boldsymbol r} {\boldsymbol r}' \rangle \rangle$, we have
\begin{eqnarray}
I_1 &=&  \langle  \Phi^\dagger_{\boldsymbol r}  \Phi^{}_{{\boldsymbol r}'} \rangle 
 \nonumber \\
&=& \frac{1}{2N_{u.c.}} \sum_{\boldsymbol k} e^{i {\boldsymbol k}\cdot {\boldsymbol a}_n } \tilde{J}_x
 \big[ \frac{1}{\omega_+ ({\boldsymbol k})} + \frac{1}{\omega_- ({\boldsymbol k})}    \big],
\end{eqnarray}
where ${\boldsymbol a}_n$ is the vector connecting ${\boldsymbol r}'$
 and ${\boldsymbol r}$. Due to the spatial uniformity, $I_1$ is uniform and real. 
 For the first-neighbor sites $\langle {\boldsymbol r} {\boldsymbol r}'  \rangle$, we have
 \begin{eqnarray}
I_2 &=&  \langle  \Phi^\dagger_{\boldsymbol r}  \Phi^{}_{{\boldsymbol r}'} \rangle 
 \nonumber \\
&=& \frac{1}{2N_{u.c.}} \sum_{\boldsymbol k} e^{-i {\boldsymbol k}\cdot {\boldsymbol e}_{\nu} }      
\frac{\sum_{\mu}   e^{i {\boldsymbol k}\cdot {\boldsymbol e}_{\mu} } }{|\sum_{\mu}   e^{i {\boldsymbol k}\cdot {\boldsymbol e}_{\mu} } |} \tilde{J}_x
 \big[ \frac{1}{\omega_- ({\boldsymbol k})} - \frac{1}{\omega_+ ({\boldsymbol k})}    \big],
 \nonumber \\
\end{eqnarray}
where ${\boldsymbol e}_{\nu}$ is the vector connecting ${\boldsymbol r}'$
 and ${\boldsymbol r}$. Here $I_2$ describes the inter-sublattice hopping/hybridization
 on the diamond lattice. It is zero in the absence of the magnetic order, and the two spinon bands
 are degenerate. 
 We should emphasize that, in the calculation of $I_1$, $I_2$ and $\langle  \Phi_{\boldsymbol r}^\dagger \Phi^{}_{\boldsymbol r}  \rangle$,
 we are working with the gapped spinons. 
 
The parameters $I_1$ and $I_2$ are then feed back to the gauge field sector. We solve 
the spinon sector and the gauge field sector together self-consistently. The self-consistent
mean-field results are finally determined based on their variational energy $\langle H_a \rangle$
where $H_a$ is given in Eq.~7 of the main text. We evaluate $\langle H_a \rangle$ with respect to the
gauge mean-field states as
\begin{eqnarray}
\langle H_a \rangle &=& \sum_{{\boldsymbol r} } \langle  \frac{\tilde{J}_x}{2}    Q_{\boldsymbol r}^2
 \rangle \nonumber \\
&-&
  \sum_{{\boldsymbol r} } \sum_{ \mu \neq \nu}{\frac{\tilde{J}_{\perp}}{2}} 
\langle 
{ {\Phi^\dagger_{{\boldsymbol r} + \eta_{\boldsymbol r} {\boldsymbol e}_{\mu}}} 
\Phi^{}_{{\boldsymbol r}  +\eta_{\boldsymbol r} {\boldsymbol e}_{\nu}  }} 
 \tilde{s}^{-\eta_{\boldsymbol r}}_{{\boldsymbol r}, {\boldsymbol r} + {\eta}_{\boldsymbol r} {\boldsymbol e}_{\mu}} 
 \tilde{s}^{+\eta_{\boldsymbol r}}_{{\boldsymbol r},{\boldsymbol r}  +\eta_{\boldsymbol r} {\boldsymbol e}_{\nu}  }  
 \rangle
 \nonumber \\
 &-& 
  \sum_{{\boldsymbol r} \in \text{I} } \sum_{\mu} \sum_{ j \in [{\boldsymbol r},{\boldsymbol r} + {\boldsymbol e}_{\mu}]_2}  
  \frac{J_{2xz}}{2}  \langle   \big( {\Phi^\dagger_{\boldsymbol r} \Phi_{{\boldsymbol r} + {\boldsymbol e}_{\mu}}^{}  
 \tilde{s}^+_{{\boldsymbol r},{\boldsymbol r} + {\boldsymbol e}_{\mu}}  
+  h.c. } \big)
 \tilde{s}^x_j \rangle
 \nonumber \\
 &=& 
 \sum_{\boldsymbol k} \frac{1}{2}\big[ \omega_+ ({\boldsymbol k})  +  \omega_- ({\boldsymbol k}) \big]
 \nonumber \\
 && \quad\quad 
 - 12 \tilde{J}_{\perp} \chi_1 I_1 N_{u.c.}- 48 J_{2xz} I_2 \chi_2 N_{u.c.},
\end{eqnarray}
where in the evaluation of the first kinetic energy term, we have made use of the factor
that, for each harmonic oscillator, the kinetic energy of the ground state is simply 1/2 of the total 
zero-point energy. As this is a complex field, we have an extra factor of 2 for the kinetic energy part. 
The remaining contributions are simply evaluated against their mean-field values in the real space.


\section{Spinons in U(1) spin liquid}

The spinons in both pyrochlore U(1) spin liquid and the pyrochlore AF$^{\ast}$ state are fully gapped. 
The difference is that, for the pyrochlore U(1) spin liquid under consideration in Fig.~1, the spinons are degenerate,
while for the pyrochlore AF$^{\ast}$ state, the spinons are not degenerate due to the inter-sublattice hopping from the 
magnetic order. As a comparison, we plot the spinon continuum for the pyrochlore U(1) spin liquid in Fig.~\ref{fig4}.
The spinon dispersion is obtained from Eq.~11 of the main text by setting $t_2=0$. This spinon dispersion
is equivalent to $\tilde{J}_{\perp}=0.1 \tilde{J}_x$ in Fig.~1 of the main text. 
Although this is not a parallel comparison, the bandwidth of the spinon continuum is much reduced in 
the pyrochlore U(1) spin liquid.

\section{With dominant $\tilde{S}^z$ interaction}

Throughout this paper, we have considered a predominant and frustrated $\tilde{S}^x$ interaction
and analyzed the resulting magnetic orders and phases. In fact, due to the identical symmetry property,
it would be qualitatively equivalent to consider the case with a predominant and frustrated $\tilde{S}^z$ interaction.
One can still obtain the pyrochlore U(1) spin liquid, the pyrochlore AF$^{\ast}$ and the fragmented AFM states. 
The procedures are identical to the ones that have been used in this paper, and the phases have qualitatively
the same physical properties. The reason to select the predominant $\tilde{S}^x$ interaction is based on the 
Curie-Weiss temperature. For the nearest-neighbor XYZ model for the DO doublet, we have~\cite{PhysRevB.95.041106} 
\begin{eqnarray}
\Theta_{\text{CW}} = \frac{1}{2} J_z.
\label{eqcw}
\end{eqnarray}
The Curie-Weiss temperature of Nd$_2$Sn$_2$O$_7$ between 5K and 15K is $-0.32$K~\cite{PhysRevB.92.144423}. 
Even though the further neighbor interactions would modify Eq.~\eqref{eqcw}, this still give a rough estimate that
$J_z$ is more likely to be ferromagnetic. 
Thus, we consider a predominant and frustrated $\tilde{S}^x$ interaction in this paper. 
In the end, $\tilde{S}^x$ and $\tilde{S}^z$ are simply notations for calculation, only the physical $S^x$ and $S^z$
are meaningful.

\section{Spinon interactions}

In this paper, we have considered the mechanism of mutual induction between the gauge link and the gauge field as the 
mechanism for the pyrochlore AF$^{\ast}$ state. 
It is also possible that, the magnetic order is induced by the condensation of spinon bilinears in the particle-hole channel
from the spinon interaction. 
The spinon interaction is universally present in the pyrochlore U(1) spin liquid. The more significant ones are the short-range
interaction from the exchange interaction. In fact, in Ref.~\onlinecite{PhysRevLett.112.167203}, the current
authors and collaborators have shown that, the XYZ model already contains a significant spinon interaction. 
Over there, the spinon interaction was not found to generate an intermediate AF$^{\ast}$-like state between
the U(1) spin liquid and the ordered state. It is, however, still reasonable for us to vision that some other spin exchange
may stabilize the pyrochlore AF$^{\ast}$ state purely from the spinon interactions. 

Despite the possibility of other mechanisms, the pyrochlore AF$^{\ast}$ state, once it is realized, should
have the expected physical properties. It is these exotic physical properties and the related phenomena that
stand out and make the candidate system Nd$_2$Sn$_2$O$_7$ exciting.

\section{Possibility of the frustrated $\pi$-flux regime}

In this paper, we have chosen the $0$-flux for the parent pyrochlore ice U(1) spin liquid. 
Although the sign of $J_z$ is ferromagnetic, it is still possible that the rotated transverse coupling
can be frustrated. In the case, one would encounter the frustrated $\pi$ flux U(1) spin liquid.
This seems to happen in the Ce-based pyrochlore Ce$_2$Sn$_2$O$_7$ and Ce$_2$Zr$_2$O$_7$
where the candidate spin liquid was expected to U(1)$_{\pi}$ spin liquid
of the octupolar type~\cite{PhysRevResearch.2.013334,PhysRevB.95.041106,PhysRevX.12.021015,Gao_2019,Bhardwaj_2022,Sibille_2020}.

For the U(1)$_{\pi}$ spin liquid, the spinon realizes the lattice translation symmetry in a projective fashion~\cite{PhysRevB.86.104412,PhysRevB.96.085136,PhysRevLett.121.067201}, and the proximate states that grow out of it would naturally break the lattice
translation symmetry. 
For the all-in all-out magnetic order in the Nd-based pyrochlore, the magnetic unit cell is identical to 
the crystal unit cell, and thus the system preserves the lattice translation. Thus, we do not consider the 
possibility of the frustrated $\pi$-flux regime.

\section{Spin-wave theory for ``fragmented AFM''}

It seems that the other Nd-based dipole-octupole pyrochlore magnets including Nd$_2$GaSbO$_7$, 
Nd$_2$Zr$_2$O$_7$ and Nd$_2$Hf$_2$O$_7$ all develop the all-in all-out magnetic order. 
Nd$_2$Zr$_2$O$_7$ and Nd$_2$Hf$_2$O$_7$ are well-interpreted from the picture of 
moment fragmentation~\cite{Petit2016,PhysRevB.94.104430,PhysRevLett.124.097203,PhysRevX.4.011007}. 
While these two systems were shown to have the all-in all-out magnetic order, 
the inelastic neutron scattering pattern at low temperatures develops pinch-point-like structures
that are reminiscent of the spin ice, indicating the presence of the spin-ice-like correlation in the 
spin fluctuations. 
The ``fragmented AFM'' in Fig.~1 of the main text and Fig.~\ref{fig3} turns to belong to the same phase
as the ones that were discussed in Nd$_2$Zr$_2$O$_7$ and Nd$_2$Hf$_2$O$_7$. 
To capture the dynamical property of this phase, it is illuminating to compute the magnetic excitation 
for the fragmented AFM and compare with the spinon continuum for the U(1) spin liquid and 
the AF$^{\ast}$ state.  

As we have remarked, the fragmented AFM develops an experimentally visible order in the $S^z$ component,
and an invisible order in the $S^x$ component. What is more, $S^z$ could also function as a disordering
operator to quantum mechanically flip the $S^x$ order and generate the magnetic excitation. 
That is precisely the reason why one can observe the magnetic excitation with the inelastic 
neutron scattering measurement.

\begin{figure}[t] 
	\centering
	\includegraphics[width=8cm]{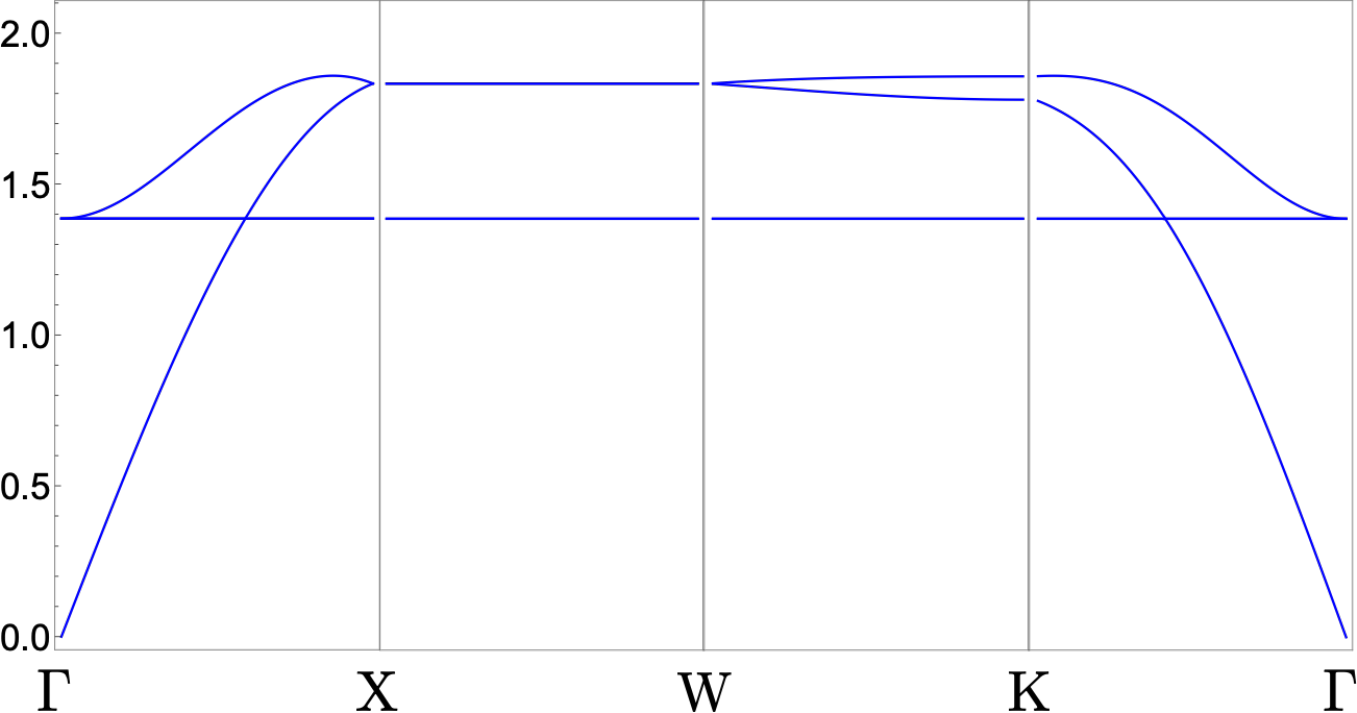}
 	\caption{
The spin-wave dispersion along the high symmetry momentum lines 
for the fragmented AFM state. In the plot, we have set 	
 	$\tilde{J}_{\perp} = 0.6$, and $\tilde{J}_x$ is set to unity. 
 	Two flat bands are degenerate. 
	}
	\label{fig5}
\end{figure}

To compute the magnetic excitation, we stay on the horizontal axis of Fig.~1 of the main text and consider the model with
\begin{eqnarray}
H= \sum_{\langle ij \rangle} \tilde{J}_x \tilde{S}^x_i {\tilde S}^x_j  -  \tilde{J}_{\perp} ( {\tilde{S}}^z_i {\tilde{S}}^z_j  + S^y_i S^y_j).
\label{eq34}
\end{eqnarray}
This simplification avoids the complicated crossing coupling between $\tilde{S}^x$ and $\tilde{S}^z$ 
from the second neighbor, but still captures the fragmented nature of the ordering. 
This Hamiltonian in Eq.~\eqref{eq34} has an accidental U(1) symmetry that should be absent in the 
generic case. 
For a large $\tilde{J}_{\perp}$, the system develops magnetic order in the $yz$ plane, and we choose
the order to be on the $\tilde{S}^z$ such that the order is smoothly connected to the generic case with
a finite $J_{2xz}$. We then perform the spin wave expansion and replace the spin operator
with the Holstein-Primarkoff bosons, 
\begin{eqnarray}
\tilde{S}^z_i &=& \frac{1}{2}-a^\dagger_i a^{}_i , \\
\tilde{S}^x_i &=& \frac{1}{2} (a^\dagger_i + a^{}_i) , \\
S^y_i &=& \frac{1}{2i} (a^{}_i - a^\dagger_i) ,
\end{eqnarray}
and the Hamiltonian in the linear spin wave theory becomes
\begin{eqnarray}
H &=& \sum_{\langle ij \rangle}  \frac{\tilde{J}_{\perp}}{2} (a^\dagger_i a_i^{} + a^\dagger_j a^{}_j) +  \frac{\tilde{J}_x + \tilde{J}_{\perp}}{4} (a_i a_j + a^\dagger_i a^\dagger_j) 
\nonumber \\
&&\quad\quad  + \frac{\tilde{J}_x - \tilde{J}_{\perp}}{4} (a_i^\dagger a^{}_j + a_j^\dagger a^{}_i ) + const,
\end{eqnarray}
where the `const' is related to the classical energy and does not influence the magnetic excitation. 
The magnetic excitations are depicted in Fig.~\ref{fig5}. There exist two flat bands 
that arise from the frustrated interaction in the $\tilde{S}^x$ component. 
The spin-ice-like 
correlation was also argued to be originated from the frustrated $\tilde{S}^x$ interaction~\cite{PhysRevB.94.104430}.
In Fig.~\ref{fig5}, we have a gapless mode at $\Gamma$ point. This is a Goldstone mode due to the 
U(1) symmetry breaking of the magnetic order for our model in Eq.~\eqref{eq34}. In the generic case, 
we do not have such a continuous symmetry,
and we should always expect a gapped spin-wave spectrum.

In the inelastic neutron scattering measurement, only the $S^z$-$S^z$ correlation is directly visible. 
The $S^z$ operator, under the spin wave expansion, is given as
\begin{eqnarray}
S^z_i &=& \cos \theta \tilde{S}^z_i + \sin \theta \tilde{S}^x_i  \nonumber \\
&=& \cos \theta (\frac{1}{2} - a^\dagger_i a^{}_i) +\frac{1}{2} \sin \theta (a^{\dagger}_i + a^{}_i) ,
\label{eq39}
\end{eqnarray}
where the ordered piece simply contributes to the magnetic Bragg peak, and the second term 
in Eq.~\eqref{eq39} generates the magnetic excitations with single magnons. 
Moreover, the $a^\dagger_i a^{}_i$ in Eq.~\eqref{eq39} would generate a two-magnon continuum
that should appear at high energies that the ones in Fig.~\ref{fig5}. This two-magnon continuum
might have a very low intensity but seems unavoidable. So far, this two-magnon continuum has not
been emphasized in the literature of the Nd-based pyrochlore magnets.

 \end{document}